# Observing Nearby Nuclei on Paramagnetic Trityls and MOFs via DNP and Electron Decoupling

Kong Ooi Tan[a,b,c], Luming Yang[a,e], Michael Mardini[a,b], Choon Boon Cheong[a,b,f], Benoit Driesschaert[d], Mircea Dincă[a], and Robert G. Griffin[a,b,*]

[a]Department of Chemistry, Massachusetts Institute of Technology, Cambridge, MA 02139, USA

[b]Francis Bitter Magnet Laboratory, Massachusetts Institute of Technology, Cambridge, MA 02139, USA

[c]Laboratoire des Biomolécules, LBM, Département de Chimie, École Normale Supérieure, PSL University, Sorbonne Université, CNRS, 75005 Paris, France

[d]Department of Pharmaceutical Sciences, School of Pharmacy, West Virginia University, Morgantown, WV, 26506 USA

[e] Research Group EPR Spectroscopy, Max Planck Institute for Multidisciplinary Sciences, Göttingen 37077, Germany

[f] Institute of Sustainability for Chemicals, Energy and Environment, 1 Pesek Road, Jurong Island, Singapore 627833, Singapore

*Corresponding Author:

Robert G. Griffin
Francis Bitter Magnet Laboratory and Department of Chemistry
Massachusetts Institute of Technology
Cambridge, Massachusetts 02139, USA
E-Mail: rgg@mit.edu

**Abstract**

Dynamic nuclear polarization (DNP) is an NMR sensitivity enhancement technique that mediates polarization transfer from unpaired electrons to NMR-active nuclei. Despite its success in elucidating important structural information on biological and inorganic materials, the detailed polarization-transfer pathway—from the electrons to the nearby and then the bulk solvent nuclei, and finally to the molecules of interest—remains unclear. In particular, the nuclei in the paramagnetic polarizing agent play significant roles in relaying the enhanced NMR polarizations to more remote nuclei. Despite their importance, the direct NMR observation of these nuclei is challenging because of poor sensitivity. Here, we show that a combined DNP and electron decoupling approach can facilitate direct NMR detection of these nuclei. We achieved an ~80 % improvement in NMR intensity via electron decoupling at 0.35 T and 80 K on trityl radicals. Moreover, we recorded a DNP enhancement factor of $\varepsilon \sim 90$ and ~11 % higher NMR intensity using electron decoupling on paramagnetic metal-organic framework, magnesium hexaoxytriphenylene (MgHOTP MOF).

# 1. Introduction

Dynamic nuclear polarization (DNP) is an NMR hyperpolarization technique that mediates polarization transfer from unpaired electrons to NMR-active nuclei via microwave irradiations. For an ideal two-spin electron-$^1$H spin system, the maximum theoretical $^1$H enhancement factor can reach $\varepsilon \sim 658$.[1–3] The method has allowed important structural information to be extracted from small molecules, biological samples, and inorganic materials.[4–6] Despite many successful applications, there is no detailed understanding of how the large electron polarization is transferred to the surrounding nuclei, or where these nuclei are located relative to the polarizing agent. Subsequently, these "nearby nuclei" are important in mediating transfer of the enhanced NMR polarizations to the target molecules via the spin diffusion mechanism.[7–10] Nevertheless, not all nearby nuclei contribute to the NMR signals observed in standard DNP experiments. Although the nearby nuclei are preferentially hyperpolarized in DNP due to the larger hyperfine interactions, their enhanced polarization might not propagate to the bulk nuclei if the spin diffusion mechanism is quenched. For example, these nuclei could have excessively broadened or shifted NMR lines and/or short $T_1$ relaxation times. Consequently, they will not be efficiently in contact with the bulk nuclei.[11,12] The region in which the hyperpolarized nuclei cannot efficiently participate in spin diffusion with the bulk nuclei is known as the 'spin diffusion barrier'.[8,11–16]

Motivated by the earlier studies, we have recently shown, using the three-spin solid effect, that the size of the spin diffusion barrier surrounding the trityl radical in a glassy glycerol-water matrix is < 6 Å.[17] More recent experimental findings have also reported similar values for various radicals under different DNP conditions.[18,19] Nevertheless, these conclusions are derived from indirect experimental results, and further information about these near nuclei—for instance, their NMR linewidths and relaxation rates ($T_1$ and $T_2$)—remains inaccessible using the standard DNP or NMR techniques. Hence, we hypothesize that a combined DNP and electron decoupling approach using two different *pulsed* microwave sources (see Section 2.2) can enable direct NMR observations of these nearby nuclear spins. In this manuscript, we define nearby nuclei as any intramolecular nuclei residing on the paramagnetic molecules with a distance of ≤ 1 nm away from the unpaired electrons. The nearby nuclei could be either inside or outside the spin diffusion barrier.

Our 0.35 T/ 15 MHz/9.8 GHz DNP spectrometer is equipped with two microwave synthesizers. The microwave frequency of the first source can be set to facilitate solid effect (SE) DNP, where the microwave (µw) offset frequency is matched to the nuclear Larmor frequency $(\Omega_{\mu w1} = \omega_{0I})$ during the polarization buildup time (**Fig 1**).[20] Additionally, the second microwave source is configured to perform electron decoupling, i.e., nutating EPR lines $(\Omega_{\mu w2} = 0)$, during the NMR detection period. Alternatively, a single frequency-agile microwave device that allows fast (microseconds) switching of µw frequencies can be employed. The first experimental evidence of the electron decoupling effect was demonstrated by Saliba et al. using such a device at 7 T. [21–23]

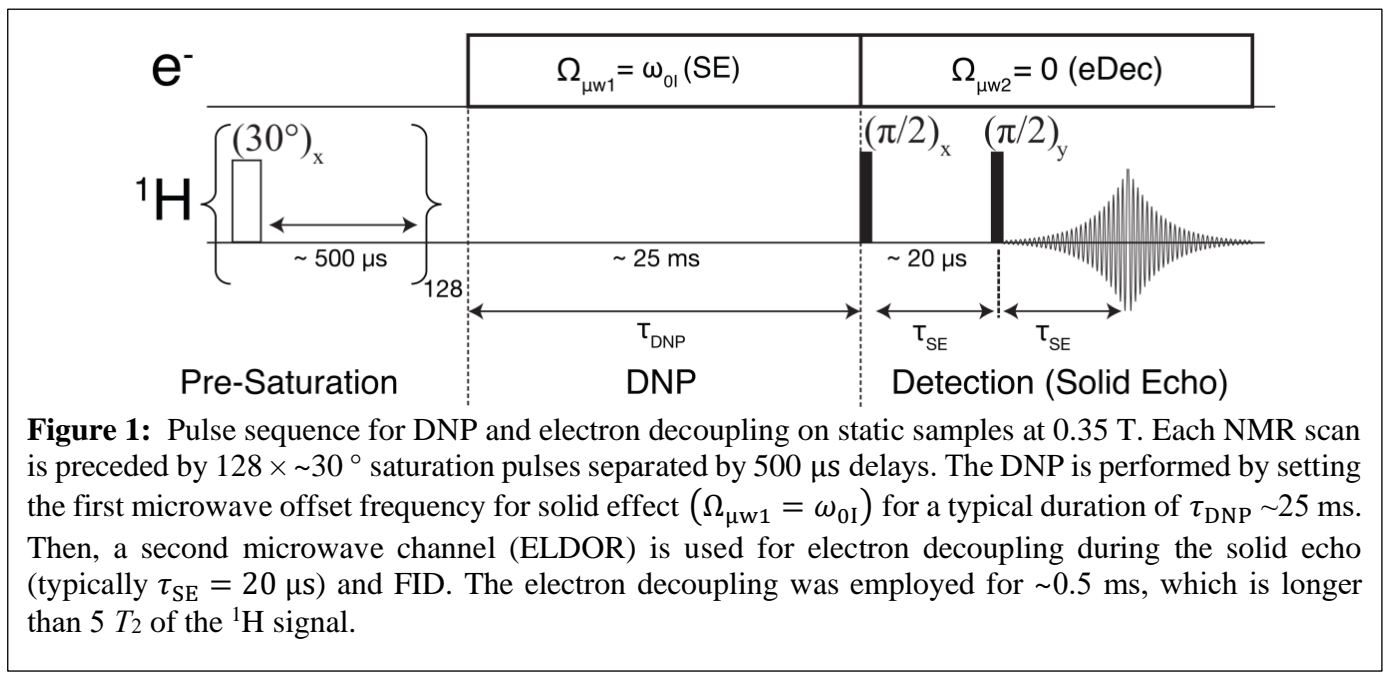

**Figure 1:** Pulse sequence for DNP and electron decoupling on static samples at 0.35 T. Each NMR scan is preceded by 128 × ~30 ° saturation pulses separated by 500 µs delays. The DNP is performed by setting the first microwave offset frequency for solid effect $(\Omega_{\mu w1} = \omega_{0I})$ for a typical duration of $\tau_{DNP}$ ~25 ms. Then, a second microwave channel (ELDOR) is used for electron decoupling during the solid echo (typically $\tau_{SE} = 20$ µs) and FID. The electron decoupling was employed for ~0.5 ms, which is longer than 5 $T_2$ of the $^1$H signal.

We expect that the study of nearby nuclei could not only shed light on spin diffusion and the fundamental DNP mechanism, but also help elucidate structural information of natural paramagnetic sites in functional biological molecules or inorganic materials, e.g., metalloproteins or battery materials etc.[24,25] We will apply this methodology to study the metal-organic framework, magnesium 2,3,6,7,10,11-hexaoxytriphenylene (MgHOTP MOF), which was recently demonstrated to be capable of quantum sensing

Li$^+$ ions using semiquinone-type radicals.[26] As the MOF itself is naturally paramagnetic, we show that the $^1$H NMR signals can be DNP-enhanced by $\varepsilon \sim 90$, followed by an additional ~11 % improvement in sensitivity via electron decoupling.

## 2. Experimental Methods

### 2.1. Sample Preparation

In contrast to procedures in standard DNP experiments, we have not used $d_8$-glycerol/D$_2$O/H$_2$O (in a 6:3:1 ratio by volume) formulation because the deuteration factor of the commercially available $d_8$-glycerol (Cambridge Isotope Laboratories, 99.5 % deuterated) is not sufficiently high; that is, any residual solvent $^1$H can overshadow the signal arises from the intramolecular $^1$H on trityls. Thus, we turned to the mixture $d_6$-DMSO/D$_2$O mixture in a 6:4 ratio (by volume) which forms glassy matrices suitable for DNP. Despite the high deuteration factor (99.96 %) of these solvents, the chemical purity of $d_6$-DMSO is only 99.5 %. Hence, any chemical impurities, if protonated, might complicate our attempts to directly observe the intramolecular $^1$H. Furthermore, a small amount of $d_8$-glycerol was added because we observed an unusual EPR lineshape and the nutation curve suggesting that the Finland trityls are more prone to aggregation in pure $d_6$-DMSO/D$_2$O than in glycerol/H$_2$O. This observation agrees with our previous finding that the trityl and glycerol preferentially associate over other solvent components.[17] Hence, we prepared a 5 mM Finland trityl in a $d_6$-DMSO/D$_2$O/$d_8$-glycerol mixture in a 57:38:5 ratio by volume, and this sample was used for the results presented in **Fig. 3** and **4a-c**.

To further minimize the $^1$H contamination in the sample from atmospheric water, we improved the sample preparation procedures. First, the radicals were mixed with solvents in a nitrogen-filled glove box, where the sample was syringed into a quartz EPR tube (Wilmad-LabGlass). Subsequently, the quartz tube was connected to a 3-way tap fitted with a 3D-printed adapter (**Fig. S1**)[27] enclosed with an O-ring before being transported out of the glove box. The glass tap was immediately fitted to a vacuum line, the headspace *before* the tap pumped and backfilled to remove contaminating atmospheric water from the transport. The tubes were then placed under vacuum and subsequently flame sealed. The DNP samples were kept frozen in liquid nitrogen (to avoid trityl aggregation) during the flame seal (the frozen sample was ~10 cm away from the flame). These samples were used for the experiments shown in **Fig. 4d**.

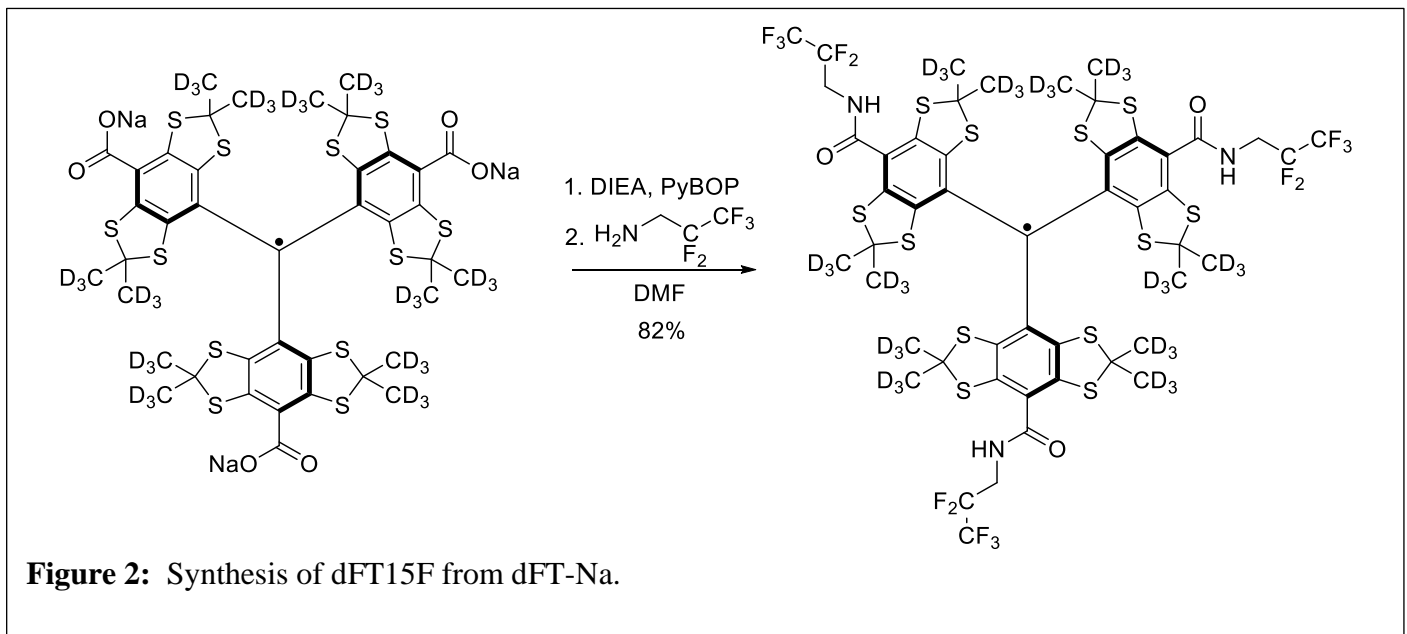

**Figure 2:** Synthesis of dFT15F from dFT-Na.

The synthesis of $^{19}$F-trityl radical, dFT15F ($C_{49}H_9D_{36}F_{15}N_3O_3S_{12}$, molecular weight: 1429.8) was performed in one step from the deuterated Finland trityl (dFT, Finland trityl-$d_{36}$) (**Fig. 2**).[28,29] Briefly, dFT sodium salt (dFT-Na) (100 mg, 0.09 mmol, 1 eq.) was dissolved in dry DMF (10 mL) under argon. N,N-Diisopropylethylamine (95 μL, 0.54 mmol, 6 eq.) was added followed by PyBOP (283 mg, 0.54 mmol, 6 eq.). The green solution immediately turned red, the solution was stirred for 5 min. Next, 2,2,3,3,3-pentafluoropropylamine (97 μL, 0.90 mmol, 10 eq.) was added, and the solution was stirred overnight. Water (10 mL) was added, then the organic layer was separated, and the aqueous layer was extracted with DCM (3 × 10 mL). The organic layers were combined, dried over MgSO$_4$, and filtered. The solvent was evaporated under reduced pressure and the crude was purified by flash chromatography (CombiFlash Rf) on silica gel (12 g) using a gradient from *n*-hexane to DCM to isolate 106 mg (82% yield) of dFT15F as a dark green solid. HRMS (ESI) m/z [M]$^+$: calculated for [$C_{49}H_9D_{36}F_{15}N_3O_3S_{12}$] 1428.2130, found 1428.2113. The $^{19}$F-trityl radical was dissolved in either deuterated tetrachloroethane (*d*-TCE) or hexafluorobenzene (HFB)-trifluoroethanol mixture (95:5 v/v) to give a 5 mM solution. For the control experiment (**Fig. 4**), OX063 trityl was dissolved in *d*-DMSO juice because the radical is insoluble in TCE.

The synthesis of MgHOTP MOF was described in the literature.[26] Note that the precursor of the MOF, i.e., the free ligand 2,3,6,7,10,11-hexahydroxytriphenylene (HHTP), is diamagnetic in typical organic solvents, but is paramagnetic when incorporated into the MOF. The radical concentration was determined to be ~30 ± 10 mM using the spin-counting technique, i.e., only ~1 – 2 % of the ligands have unpaired electrons. The radical concentration is found to be significantly lower than the expected value (~ 50 %), possibly due to antiferromagnetic spin-spin coupling mediated by the ligands.[30,31] We will not discuss further on this as the work is still ongoing and beyond the scope of this manuscript. Note that no additional radicals or solvents were added to the MOF for DNP experiments.

## 2.2. EPR and DNP NMR Spectroscopy

All EPR and DNP experiments were performed using the Bruker X-band (0.35 T) instrument described in previous publications.[17,32] The instrument was equipped with two microwave sources capable of performing pulsed electron-electron double resonance (PELDOR/DEER) or ELDOR-detected NMR experiments.[33,34] A 10 W microwave amplifier with a 100 % duty cycle was used in all experiments, and generated a Rabi field $\omega_{1S}/2\pi$ ~ 2 MHz. In the experiments described here, we used the microwave pulse forming unit (MPFU) for DNP and the ELDOR channel for electron decoupling. This configuration was chosen because the frequency of the latter μw channel can be directly configured in the PulseSPEL pulse program, which facilitates the acquisition of the electron decoupling profile. The RF circuitry was also improved for better sensitivity,[35] and 128 saturation pulses were employed to ensure that any enhanced NMR signal from previous scans was fully saturated. This is crucial because the signal intensity gain (≤ 1.8x) observed in electron-decoupled signal could be overshadowed by the sensitivity gain from DNP (≥ 100x). Note that the DNP enhancement factor represents only an estimated value based on previous studies performed on partially protonated DNP juice[17,35]. The actual DNP enhancement on the nearby nuclei could not be determined because the $^1$H thermal equilibrium signal was not observable despite long acquisition periods. To directly observe only the nearby nuclei, a short $\tau_{DNP}$ time ($\leq 25$ ms) and a recycle delay of $\sim$ 0.25 s was used.

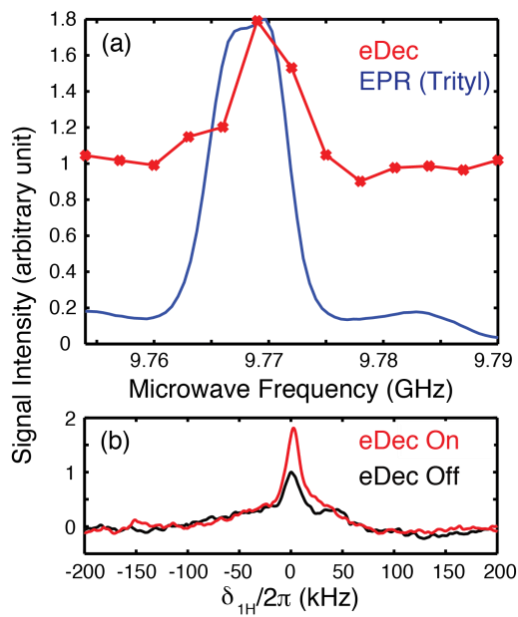

**Figure 3:** DNP and electron decoupling experiments performed on 5 mM Finland trityl sample in *d*-DNP juice at 0.35 T and 80 K. **(a)** An EPR spectrum (blue) acquired with a field-swept spin-echo sequence is shown for reference. The electron decoupling (eDec) frequency profile (red) was obtained by varying the μw frequency with a fixed 0.3484 T field. The frequency of the first μw source was set to 9.786 GHz with $\tau_{DNP} = 25$ ms to mediate the SE DNP using an offset frequency $\Omega_{\mu w1}/2\pi \sim 15$ MHz. Each NMR spectrum was recorded with an average of 4096 scans. **(b)** 1D spectra acquired at the $\Omega_{\mu w2} = 0$ condition with the second μw source used for electron decoupling turned on (red) or off (black). We have used $\tau_{SE} = 10$ μs in these experiments. Note that switching microwave sources between the DNP and electron decoupling period can be performed in $\leq 5$ ns.

## 3. Results and Discussions

**Organic Radicals**

A conventional solid-effect (SE) DNP experiment was applied to a Finland trityl sample dispersed in *d*-DNP juice using microwaves applied at the offset frequency $\Omega_{\mu w1} = \omega_{0I}$ at 0.35 T.[17,32,36–38] Subsequently, the first μw source was turned off and the second μw channel with a preset frequency ($\Omega_{\mu w2}=0$) was turned on (**Fig. 1**) to facilitate electron decoupling (eDec) during the NMR acquisition. Note that eDec was applied throughout the solid echo and the FID period. We expected that the electron-decoupled $^1$H-NMR spectrum would have a maximum integrated intensity (less signal decay during solid echo) and narrowest lines
8

(slower signal dephasing during FID) if efficient eDec is achieved. Indeed, the eDec frequency profile (**Fig. 3a**) obtained by measuring the integrated intensity as a function of µw frequency showed that a ~80 % higher signal is achieved when the µw source is on-resonance ($\Omega_{\mu w2} = 0$) with the EPR line at ~9.771 GHz. Our results are in good agreement with those reported previously by the Barnes and coworkers,[21–23] despite significantly different experimental conditions. In particular, they measured $^{13}$C-detected spectrum decoupled from electrons using frequency-chirped µw pulses with a gyrotron under magic-angle spinning at 7 T fields. We also noted that the eDec performance depends on the $\tau_{DNP}$ used in the polarization period. If shorter $\tau_{DNP}$ values are employed, then more strongly coupled nuclei can be polarized and display a higher gain when electron decoupled. However, the NMR signal would be too weak to be observed if $\tau_{DNP}$ is too short. Hence, we chose an intermediate value of $\tau_{DNP} = 25$ ms. Nevertheless, we emphasize that it is not our goal to pursue a condition (e.g., short $\tau_{DNP}$) in which the eDec is most efficient, but rather our goal is to differentiate the intramolecular nuclei from those in the solvent.

We also compared the 1D spectra (**Fig. 3b**) with and without eDec. Although a more intense line was observed in the electron-decoupled spectrum, the normalized spectra (data not shown) showed little difference. This implies that the observed peaks are still predominantly broadened by the $^1$H homonuclear dipolar couplings, and we expect that the eDec effect will be more pronounced during the solid echo period during which the $^1$H–$^1$H couplings are attenuated. To confirm this, we measured the $T_2$ relaxation times (**Fig. 4a**) under eDec by varying the solid − echo period ($\tau_{SE}$). Although an effective $T_2$ (referred to as $T_2$') is measured here instead of the actual stochastic-process-driven $T_2$, the ' sign is omitted to streamline the notation. Contrary to our expectations, although eDec resulted in overall higher line intensities, the $^1$H's coherence lifetimes appear to be shortened, i.e., the $T_2$ value decreases from 31 to 18 µs if it is electron decoupled. This anomaly could be related to the presence of some outliers in the mono-exponential fit, which implied that multiple NMR peak components might be present underneath the broad NMR spectrum. Thus, we inspected the 1D spectra and noted that the $\tau_{SE} = 10$ µs spectrum is significantly broader (**Fig. 4b**, bottom) than that acquired with $\tau_{SE} = 55$ µs. This confirms that multiple components are indeed present within the poorly resolved $^1$H peak, and the overall relaxation cannot be quantified by a single $T_2$ value. Additionally, we noted a ~100 kHz broad component that appears only in the $\tau_{SE}$ = 10 µs (blue) spectrum, and the difference between the two (black) spectra revealed a Pake-like pattern that could encode rich electron-nuclear distance information. It is evident that eDec has revealed strongly coupled, and hence severely broadened, $^1$H peaks that are otherwise not visible under conventional NMR or DNP spectroscopy. We expect these methods could be extended to determine the sizes of hyperfine interactions that contain important structural information when pulsed DNP under MAS at high field becomes feasible in the future.[39] To complete the analysis, we also determined the $T_1$ of nearby nuclei to be ~11 s, and the curve is well fitted with a mono-exponential function (**Fig. 4c**) without significant outliers.

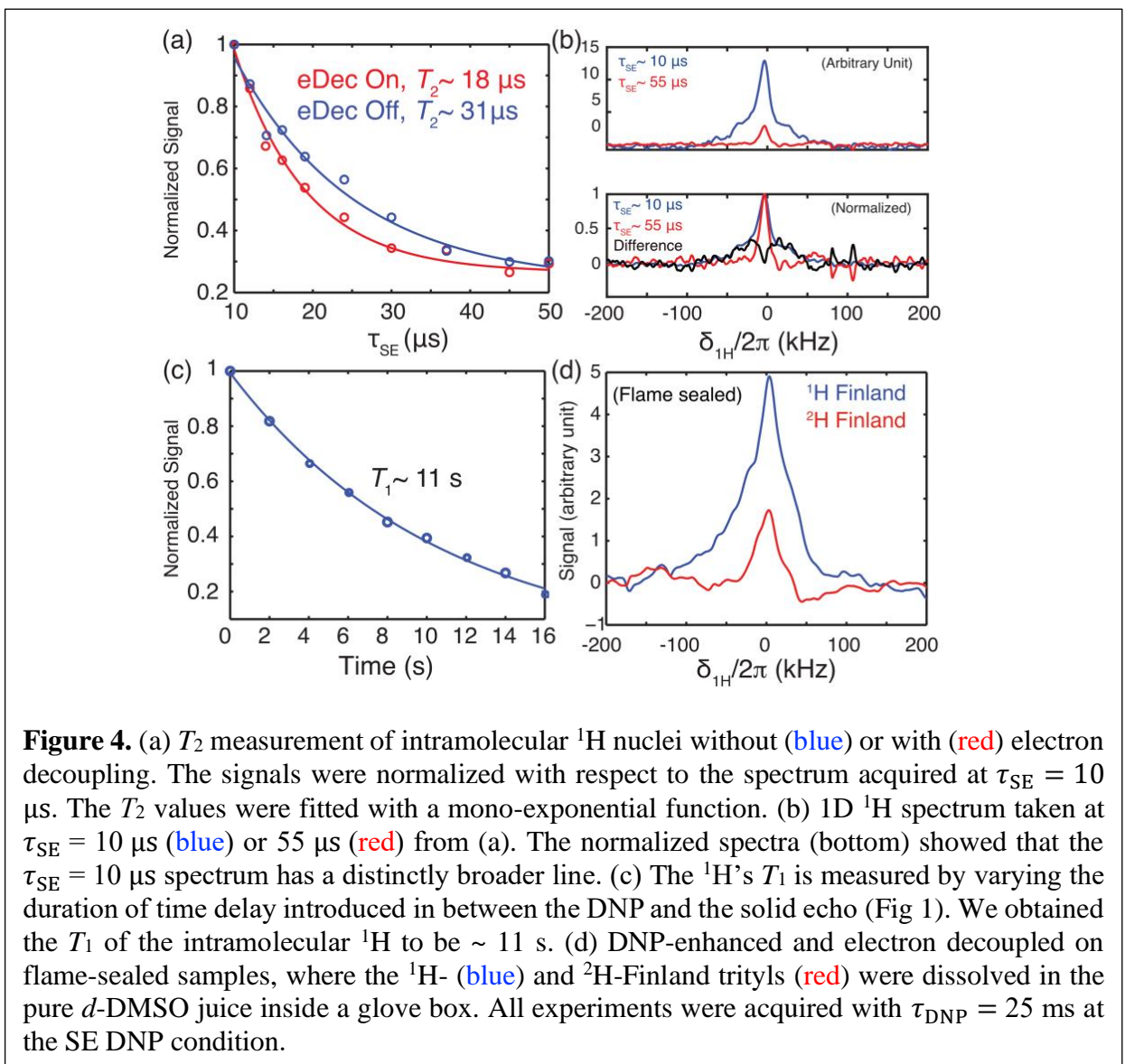

**Figure 4.** (a) $T_2$ measurement of intramolecular $^1$H nuclei without (blue) or with (red) electron decoupling. The signals were normalized with respect to the spectrum acquired at $\tau_{SE}$ = 10 µs. The $T_2$ values were fitted with a mono-exponential function. (b) 1D $^1$H spectrum taken at $\tau_{SE}$ = 10 µs (blue) or 55 µs (red) from (a). The normalized spectra (bottom) showed that the $\tau_{SE}$ = 10 µs spectrum has a distinctly broader line. (c) The $^1$H's $T_1$ is measured by varying the duration of time delay introduced in between the DNP and the solid echo (Fig 1). We obtained the $T_1$ of the intramolecular $^1$H to be ~ 11 s. (d) DNP-enhanced and electron decoupled on flame-sealed samples, where the $^1$H- (blue) and $^2$H-Finland trityls (red) were dissolved in the pure $d$-DMSO juice inside a glove box. All experiments were acquired with $\tau_{DNP}$ = 25 ms at the SE DNP condition.

We had planned to assign the ~100 kHz broader components to the intramolecular $^1$H's on trityl. However, the assignment is not unambiguous because there is still too much residual solvent $^1$H, possibly originating from the atmospheric moisture. Hence, we prepared new flame-sealed samples of either $^1$H-Finland or $^2$H-Finland trityl dissolved in the near '100 %' perdeuterated DMSO juice (see Section 2.1). Despite our best attempt to unambiguously assign the more intense $^1$H peak in the $^1$H-Finland sample (**Fig. 4d**) to the trityl $^1$H, there are still some $^1$H signals in the $^2$H-Finland sample. We attributed the unknown $^1$H source to the 0.5 % potentially protonated chemical impurity in the $d_6$-DMSO solvent.

Having realized that preparing a completely $^1$H-free sample for control is very challenging, we turned to probing intramolecular $^{19}$F in fluorinated trityl radicals, dFT15F.[28] Since $^{19}$F has a similar gyromagnetic ratio as $^1$H ($\gamma_{19F}/\gamma_{1H}$ ~0.94), the electron-nuclear and homonuclear dipolar couplings for the same distance would be slightly scaled to ~0.94 and 0.88, respectively. Given the similar strengths of dipolar interactions that drive DNP and spin diffusion, we hypothesized that the concept of spin diffusion barrier is still fundamentally applicable for the $^{19}$F system. Moreover, it is easy to adapt our experiments from $^1$H to mixture) using similar experimental conditions. The results confirm that dFT15F remains an efficient DNP polarizing agent compared to Finland or OX063 trityls.

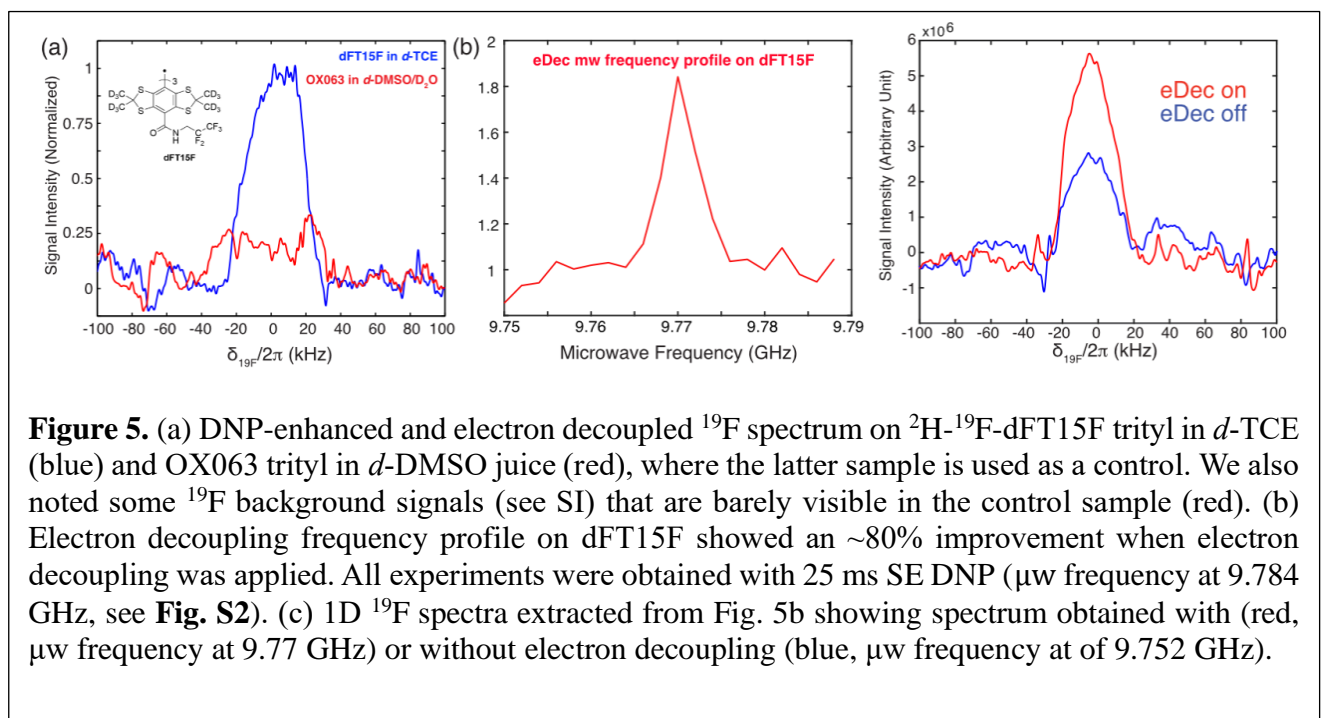

**Figure 5.** (a) DNP-enhanced and electron decoupled $^{19}$F spectrum on $^2$H-$^{19}$F-dFT15F trityl in $d$-TCE (blue) and OX063 trityl in $d$-DMSO juice (red), where the latter sample is used as a control. We also noted some $^{19}$F background signals (see SI) that are barely visible in the control sample (red). (b) Electron decoupling frequency profile on dFT15F showed an ~80% improvement when electron decoupling was applied. All experiments were obtained with 25 ms SE DNP (µw frequency at 9.784 GHz, see **Fig. S2**). (c) 1D $^{19}$F spectra extracted from Fig. 5b showing spectrum obtained with (red, µw frequency at 9.77 GHz) or without electron decoupling (blue, µw frequency at of 9.752 GHz).

$^{19}$F NMR because the difference in their Larmor frequencies is only 0.8 MHz at ~0.35 T. We have recorded a $^{19}$F DNP field profile (**Fig. S2**) of $^2$H-$^{19}$F-trityl (dFT15F) in fluorinated solvent (HFB-trifluoroethanol

We repeated the eDec experiments on dFT15F in $^{19}$F-free $d_2$-TCE, and the results (**Fig. 5a**) allow us to confidently assert that the intramolecular $^{19}$F on the trityl molecules have been directly observed. The fact that their peaks can be directly observed and not excessively broadened (linewidths of ~40 kHz) indicate that the intramolecular $^{19}$F nuclei are outside the spin diffusion barrier and, thus, contribute to the overall bulk $^{19}$F signal. This is not a surprising result as our DFT calculations show that the mean e$^-$-$^{19}$F distance is ~7.6 Å (see Table S1 in SI), which is clearly outside the spin diffusion barrier determined in previous studies.[17] Additionally, a similar ~80 % signal improvement (**Fig. 5b** and **c**) was achieved during eDec, showing again that the intramolecular nuclei on the radical can be efficiently decoupled from the unpaired electron. Following that, further characterization of the intramolecular $^{19}$F nuclei is anticipated in future work, such as to study their structures, their roles in DNP, and spin diffusion to the bulk.

**Metal Oxide Frameworks**

Following the DNP and eDec study on intramolecular nuclei on the organic radicals, we applied the same techniques on other paramagnetic materials whose hyperfine-coupled nuclei might play important roles in material science applications. For instance, MgHOTP MOF is one of the paramagnetic materials that shows promising applications in quantum sensing of Li ions via EPR.[26] However, the exact binding mechanism of the Li ions to the MOF is not yet known, and we proposed that DNP and electron decoupling could answer this question by elucidating the structures of the native MOF. First, we note that the paramagnetic MOF has a narrow EPR linewidth (~18 MHz, **Fig. 6a**), and the spin counting technique suggests that the radical concentration is ~30 mM, which is within the range typically used for DNP applications. As expected, the MOF featured a standard SE DNP profile (**Fig. 6a**) with an enhancement of $\varepsilon \sim 90$ using short buildup times ($\tau_{1DNP} \sim 0.2$ s). A possible reason for the short DNP buildup time of this MOF–about 20 times shorter than those in trityls–are that the electrons and/or $^1$H nuclei have fast relaxation rates. Indeed, Fig. **6** (c-e) shows that the electron $T_{1e}$ is ~24 µs, and the $^1$H relaxation times are $T_1$ ~ 0.22 s and $T_2$ ~ 16 µs, respectively. The similarity between the $T_1$ and $\tau_{1DNP}$ values imply that the DNP process is in the relaxation-limited regime,[14] and any additional DNP polarization built up past $\tau_{1DNP}$ time will be limited via spin-lattice relaxation.

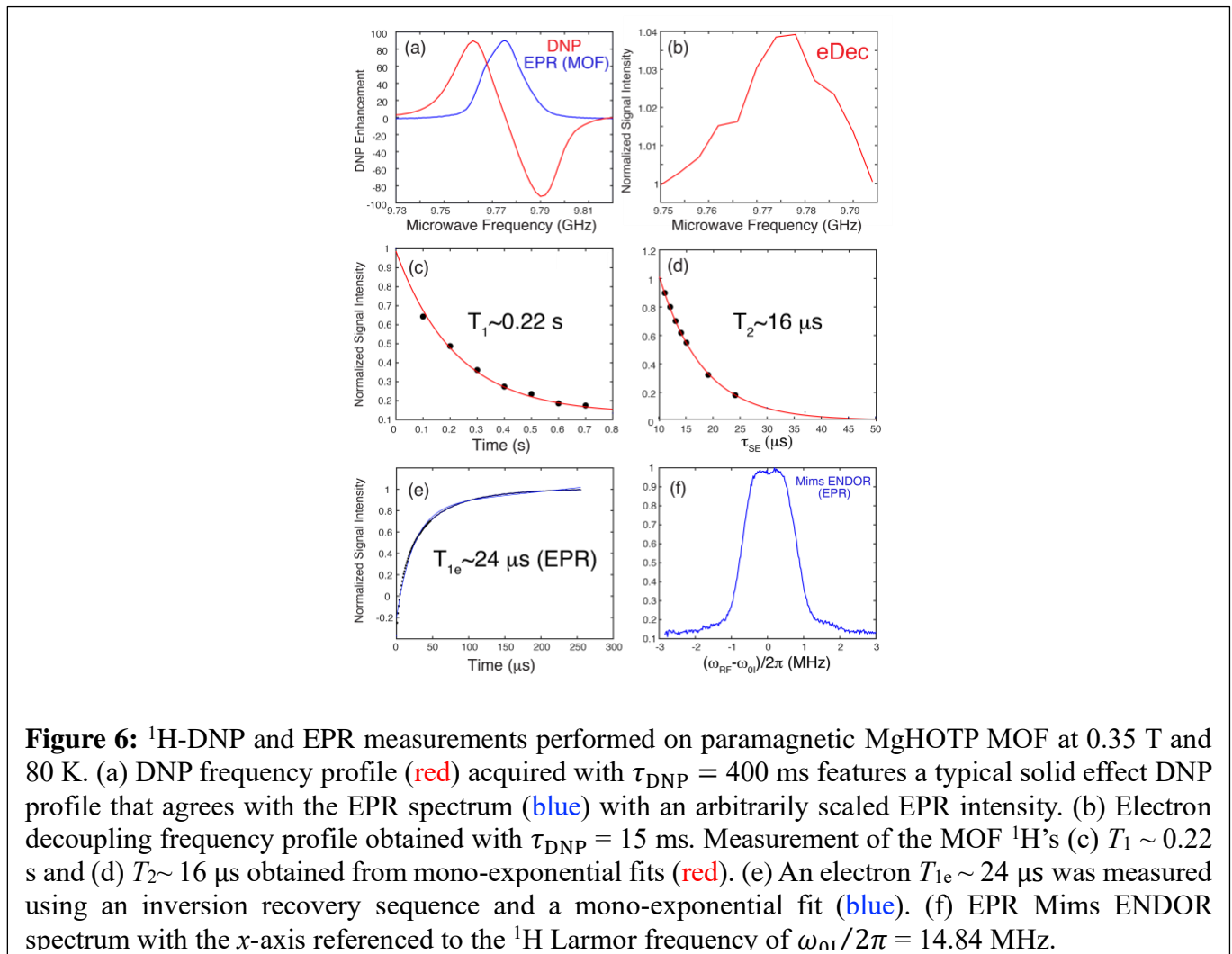

**Figure 6:** $^1$H-DNP and EPR measurements performed on paramagnetic MgHOTP MOF at 0.35 T and 80 K. (a) DNP frequency profile (red) acquired with $\tau_{DNP}$ = 400 ms features a typical solid effect DNP profile that agrees with the EPR spectrum (blue) with an arbitrarily scaled EPR intensity. (b) Electron decoupling frequency profile obtained with $\tau_{DNP}$ = 15 ms. Measurement of the MOF $^1$H's (c) $T_1$ ~ 0.22 s and (d) $T_2$~ 16 µs obtained from mono-exponential fits (red). (e) An electron $T_{1e}$ ~ 24 µs was measured using an inversion recovery sequence and a mono-exponential fit (blue). (f) EPR Mims ENDOR spectrum with the x-axis referenced to the $^1$H Larmor frequency of $\omega_{0I}/2\pi$ = 14.84 MHz.

Besides DNP, the eDec performance on $^1$H near the paramagnetic organic linker was measured to be ~4% (**Fig. 6b**), significantly lower than the ~80 % observed in trityls. The weaker decoupling performance could be due to several reasons: (1) the $^1$H's are strongly coupled in the fully protonated MOF's ligands, hence, the $^1$H's polarization are more efficiently spin-diffused to the bulk and less affected by nearby electrons, (2) the EPR linewidths in MOF are twice as broad those in trityls, which resulted in a more difficult saturation of EPR lines (shorter $T_{1e}$ and higher radical concentration), (3) stronger hyperfine interactions in MOF. To confirm the last hypothesis, a Mims ENDOR experiment was performed, and it showed a ~2 MHz broad $^1$H peak (**Fig. 6f**), which is about an order of magnitude larger than that exhibited by trityl in *d*-DNP juice.[17] Thus, we expect that the eDec performance can be significantly improved if higher microwave power is available and frequency-chirped pulses can be applied to frequency-sweep through the EPR spectrum.

## Conclusion

We have demonstrated that a combined DNP and electron decoupling approach allows an ~80% improvement in NMR signal intensity of intramolecular trityls' $^1$H and $^{19}$F nuclei. This was achieved via electron decoupling with ~2 MHz microwave Rabi field at 0.35 T. Electron decoupling has revealed some broad components (≥100 kHz) that are not easily visible in conventional DNP technique. These components might contain important distance or structural information, i.e. it could become a new method in studying paramagnetic biomolecules or materials with upcoming pulsed DNP technology.[39] We would like to emphasize that the DNP buildup time required for observing *only* the nearby nuclei is in the order of $T_{1e}$ (μs-ms) because the slower nuclear-nuclear spin diffusion process is less relevant in this context. Thus, high-sensitivity nearby nuclei spectroscopy can already be envisaged at high-field DNP with short $\tau_{1DNP}$. Furthermore, direct detection of intramolecular trityl $^{19}$F nuclei facilitated by DNP and eDec affirms that these nearby nuclei could participate in spin diffusion with the bulk nuclei during DNP. Finally, we extended the methodology to MgHOTP MOF, and the high DNP performance ($\varepsilon$ ~90) suggests that the MOF could be used for promising potential quantum sensing applications via DNP NMR.

**Acknowledgements**

We acknowledge Prof. Daniel Suess and Suppachai Srisantitham for using their glove box. We also thank Prof. J. H. Ardenkjær-Larsen (Technical University of Denmark) for providing OX063 trityl radicals. This work was supported by the National Institute of General Medical Sciences (GM132997 and GM132079). K.O.T acknowledges support from the French National Research Agency: ANR-20-ERC9-0008 and *HFPulsedDNP*, as well as RESPORE (n° 339299). B. D. acknowledges support from EB028553 and EB-023990 (synthesis of trityl radical). Work in the Dinca lab is supported by the National Science Foundation (DMR-2105495).

**Supporting Information Available**: Additional information about the 3D-printed Adapter, $^{19}$F-DNP, the HRMS spectrum and DFT calculations of the $^{19}$F trityl are available in the SI.

# Supporting Information

## 1. 3D-printed adapter for sample preparation

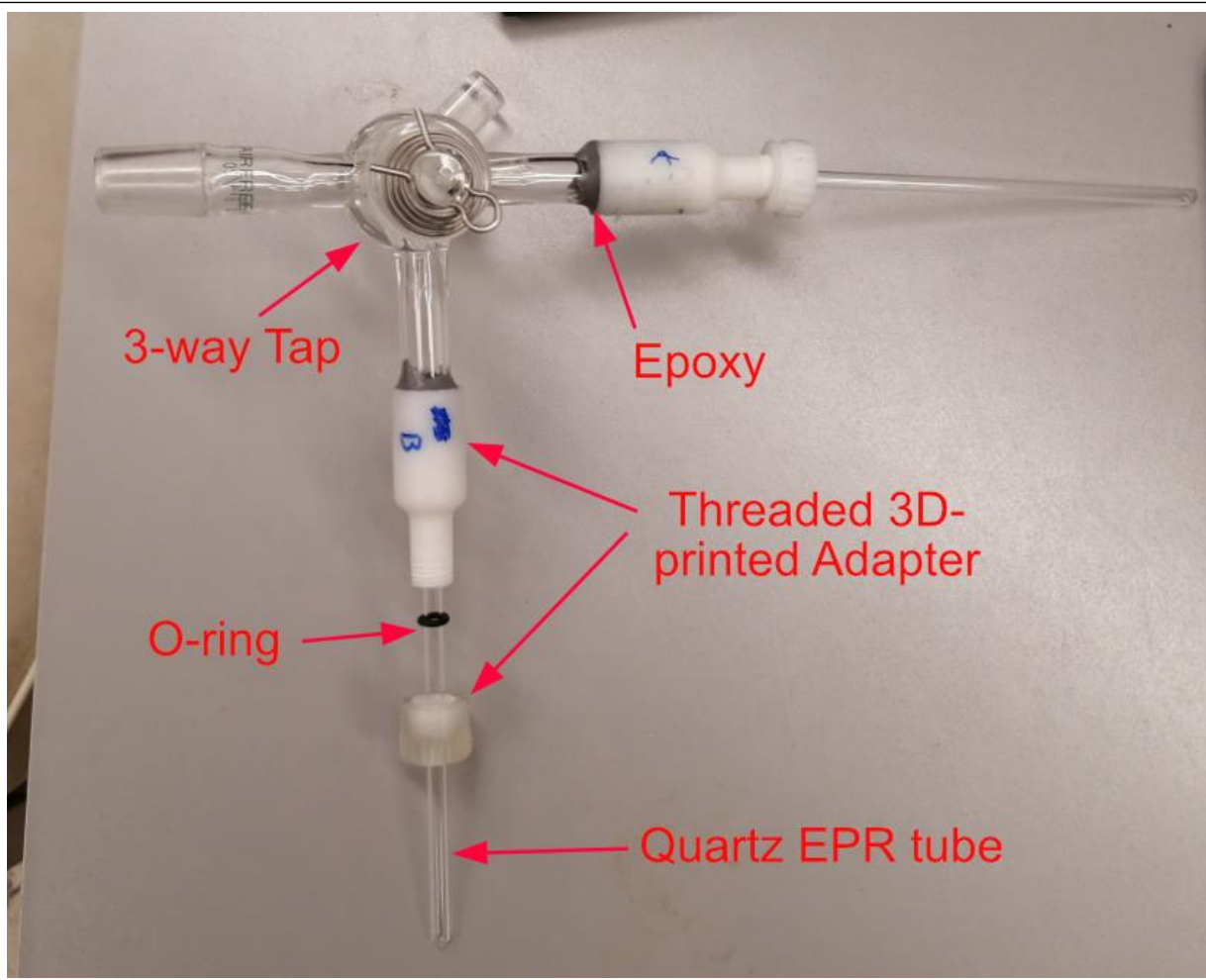

Figure S1. Picture of the homemade 3D-printed adapter printed using either clear or rigid resins with Form 3 (Formlabs Inc.). The two-piece adapter was threaded with a tap and die set to accommodate the O-ring that forms a good seal when the cap is tightened. The setup was tested to be leakproof, i.e., a minimum pressure of $10^{-2}$ mBar can be sustained while being pumped by a rotary vane pump. The adapter allows two DNP samples to be prepared in a single session.

## 2. $^{19}$F-DNP field profile on $^2$H-$^{19}$F Trityls

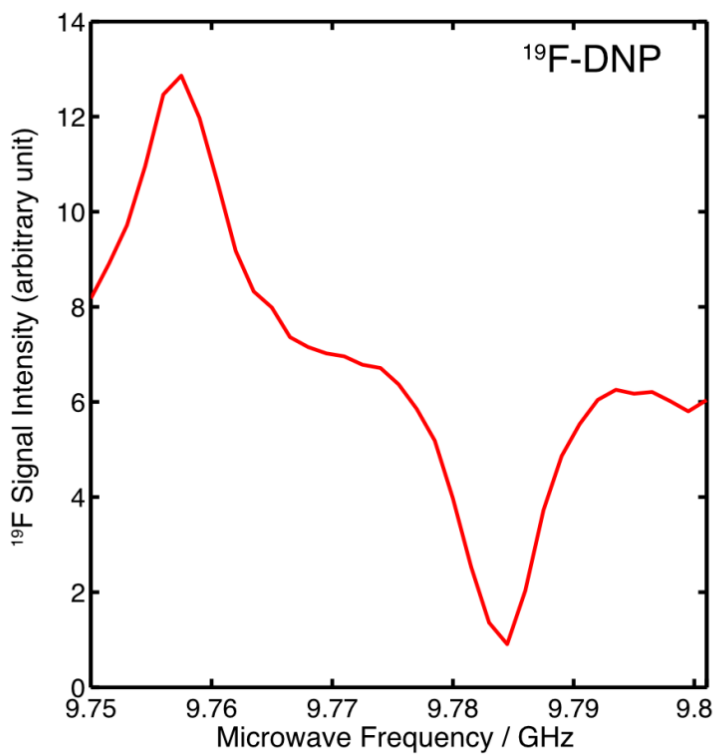

Figure S2. $^{19}$F DNP field profile of dFT15F in HFB-trifluoroethanol mixture at 80 K and 0.35 T. As expected, we observed two major peaks separated by $2\omega_{0I}/2\pi \sim 28$ MHz, which is characteristic for solid effect. The EPR spectrum (data not shown) is similar to OX063 or Finland trityl. We noticed a significant $^{19}$F background in the probe (verified by $^{19}$F NMR measurement without sample inserted in the probe). Additionally, the solvent is fully fluorinated, which contrasts with only ~10% $^1$H content in the DNP juice. Thus, we did not determine the $^{19}$F DNP enhancement because it is less relevant for electron decoupling experiments.

## 3. HRMS spectra of dFT15F

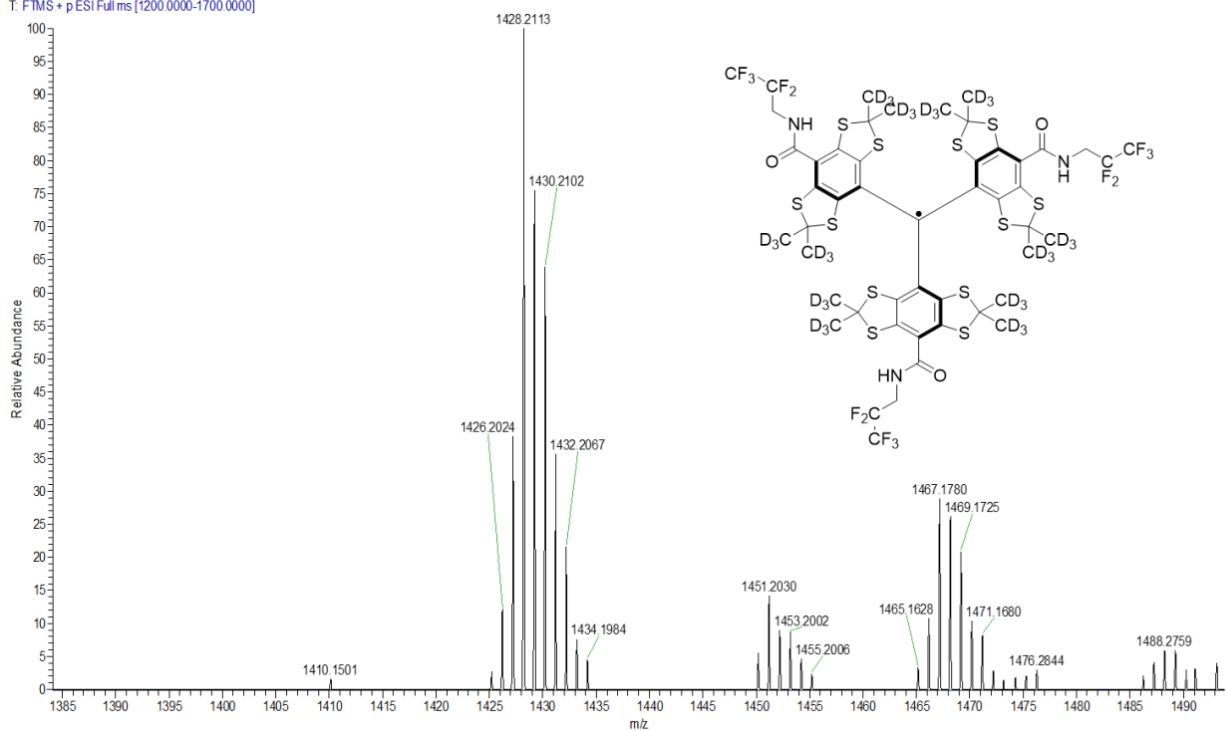

Figure S3. HRMS (ESI pos) spectrum of dFT15F.

## 4. DFT calculations of FT15F

ORCA was used for DFT calculations as previously described in Tan et. al., (doi:10.1126/sciadv.aax2743). Geometry optimization was done using the B3LYP functional and 6-31G basis set. Calculations of EPR parameters, namely the electron $g$-tensor and hyperfine couplings to all $^1$H, $^2$H, and $^{19}$F nuclei, were done with the final optimized structure and same functional with the IGLO-III basis set. The e$^-$-$^{19}$F distances are extracted by first computing the eigenvalues (principal components) of each hyperfine tensor, then subtracting the isotropic component, leaving the diagonal elements of the traceless anisotropic hyperfine interaction that are then used to compute the electron-nuclear distance. We note that some anisotropic hyperfine interactions do not have diagonal elements in the form of [-1, -1, 2] in the principal-axis system, i.e., a point-dipole approximation is not strictly valid in these cases. The largest deviation observed for e-$^{19}$F dipolar coupling has diagonal elements of [-0.65, -1.35, 2]. Nevertheless, most e-$^{19}$F couplings have diagonal elements approximately close to [-1, -1, 2]. We have tabulated the statistics concerning the e-$^{19}$F of dFT15F in Table S1.

| Group | Mean Distance | Standard Deviation | Minimum | Maximum |
|---|---|---|---|---|
| All $^{19}$F atoms | 7.6 Å | 1.1 Å | 6.0 Å | 9.1 Å |
| $CF_2$ | 6.8 Å | 0.6 Å | 6.0 Å | 7.7 Å |
| $CF_3$ | 8.2 Å | 1.0 Å | 6.1 Å | 9.1 Å |

Table S1 electron-$^{19}$F distance information determined by DFT calculations.

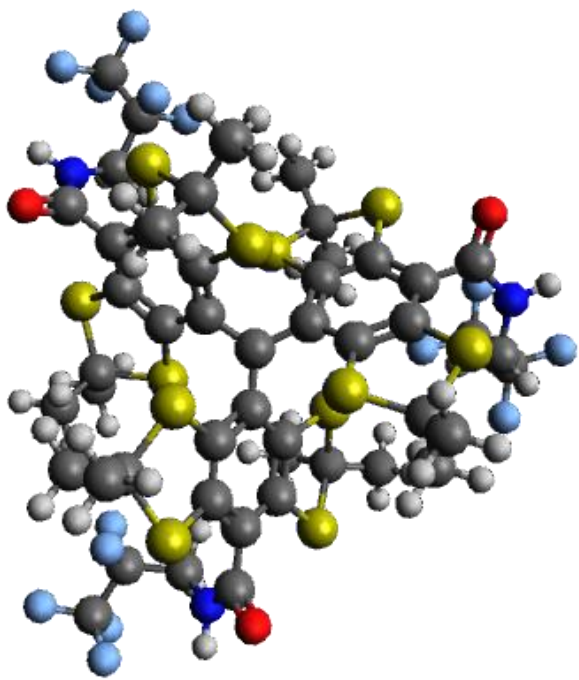

Figure S4. DFT-optimized structure of dFT15F. The atoms are colour-coded as follows: carbon (black), hydrogen (grey), nitrogen (blue), oxygen(red), and fluorine (light blue).